\documentclass{icrc2009}
\usepackage{graphicx}   
\usepackage{caption}    
\usepackage[font=footnotesize]{subfig} 
\usepackage{fixltx2e}
\usepackage{url}

\newcommand{\shorttitle}[1]%
{\markboth{Proceedings of the 31\MakeLowercase{$^{st}$} ICRC, {\L}\'{o}d\'{z} 2009}{#1} }

\begin{document}
\title{Large Scale Cosmic Rays Anisotropy With IceCube}

\author{\IEEEauthorblockN{
			   Rasha U Abbasi\IEEEauthorrefmark{1},
                           Paolo Desiati\IEEEauthorrefmark{1}
                           and
                           J. C. Diaz Velez\IEEEauthorrefmark{1} for the IceCube Collaboration \IEEEauthorrefmark{2}}
                            \\
\IEEEauthorblockA{\IEEEauthorrefmark{1} University of Wisconsin, IceCube Neutrino Observatory, Madison, WI~53703, USA}
\IEEEauthorblockA{\IEEEauthorrefmark{2} http://icecube.wisc.edu/}
}

\maketitle


\begin{abstract} 

We report on a study of the anisotropy in the arrival direction of cosmic rays with a median energy per Cosmic Ray (CR) particle of $\sim 14$ TeV using data from the IceCube detector. IceCube is a neutrino observatory at the geographical South Pole, when completed it will comprise 80 strings plus 6 additional strings for the low energy array Deep Core. The strings are deployed in the deep ice between 1,450 and 2,450 meters depth, each string containing 60 optical sensors. The data used in this analysis were collected from April 2007 to March 2008 with 22 deployed strings. The data contain $\sim 4.3$ billion downward going muon events. A two-dimensional skymap is presented with an evidence of $0.06 \%$ large scale anisotropy. The energy dependence of this anisotropy at median energies per CR particle of 12 TeV and 126 TeV is also presented in this work. This anisotropy could arise from a number of possible effects; it could further enhance the understanding of the structure of the galactic magnetic field and possible cosmic ray sources.
\end{abstract}

 

\section{Introduction}

The intensity of Galactic Cosmic Rays (GCRs) have been observed to show sidereal anisotropic variation on the order of $10^{-4}$ at energies in the range of 1-100 TeV (\cite{nag},~\cite{ta} and ~\cite{sk}). This anisotropy could arise from number of different combination of causes. One possible cause could be the Compton Getting (CG) effect. This effect was proposed in 1935~\cite{cg} predicting that CR anisotropy could arise from the movement of the solar system around the galactic center with the velocity of $\sim 220~km s^{-1}$ such that an excess of CR would be present in the direction of motion of the solar system while a deficit would appear in the opposite direction. Another possible effect (proposed by Nagashima \textit{et. al.}~\cite{nag}) causing the excess in the anisotropy, which was referred to as "tail-in", originates from close to the tail of the heliosphere.  While the deficit in the anisotropy, which was referred to as "loss-cone", originates from a magnetic cone shaped structure of the galactic field in the vicinity region. 

In this paper we present results on the observation of large scale cosmic ray anisotropy by IceCube. Previous experiments have published a 2-dimensional skymap of the northern hemisphere sky (\cite{ta}-~\cite{milagro}). This measurement presents the first 2-dimensional skymap for the southern hemisphere sky. In addition, we present the energy dependence of this anisotropy at median energies per CR particle of 12 TeV and 126 TeV. 

The outline of the paper is: the second section will describe the data used in this analysis, the analysis method and the challenges. The third section will discuss the results and the stability checks applied to the data. The fourth section will discuss the anisotropy energy dependence and the last section is the conclusion. 

\section{Data Analysis}
\label{sect-selection}

 The data used in this analysis are the downward going muons collected by the IceCube neutrino observatory comprising 22 strings. The data were collected from June 2007 to March 2008. The events used in this analysis are those reconstructed by an online one iteration Likelihood (LLH) based reconstruction algorithm.  The events selected online require at least ten triggered optical sensors on at least three strings. The average rate of these events is  $\sim240$ Hz (approximately 40 $\%$ of the events at triggering level). Further selection criteria are applied to the data to ensure good quality and stable runs. The final data set consists of $4.3 \times 10^{9}$ events with a median angular resolution (angle between the reconstructed muon and the primary particle) of $3^{\circ}$ and a median energy per CR particle of 14 TeV as simulated according to CORSIKA~\cite{sim1} using SIBYLL~\cite{sim2} hadronic interactions model and H$\ddot{o}$randel~\cite{sim3} primary cosmic ray spectrum.

In this analysis we are searching for a high precision anisotropy.  The sidereal variation of the CR intensity is induced by the anisotropy in their arrival direction. However, it can also be caused by the detector exposure asymmetries, non-uniform time coverage, diurnal and seasonal variation of the atmospheric temperature. Apart from these effects the remaining variations can only be of galactic origin.

Due to the unique location of IceCube at the South Pole the detector observes the sky uniformly. This is not the case for all the previous experiments searching for large scale cosmic ray anisotropy (e.g. ~\cite{ta}, ~\cite{sk}, ~\cite{milagro}).  Due to their locations they need a whole solar day to scan the entire sky. As a result they need to eliminate the diurnal and seasonal variations using various techniques. For IceCube the sidereal variation is not affected by diurnal variations because the whole sky is fully visible to the detector at any given time and because there is only one day and one night per year. In addition, although the seasonal variation of muon event rate is on the order of $20 \%$ the variation is slow and does not affect the daily muon intensity significantly.

\begin{figure}[h!]
 \begin{center}
 \noindent
 \includegraphics[width=6cm]{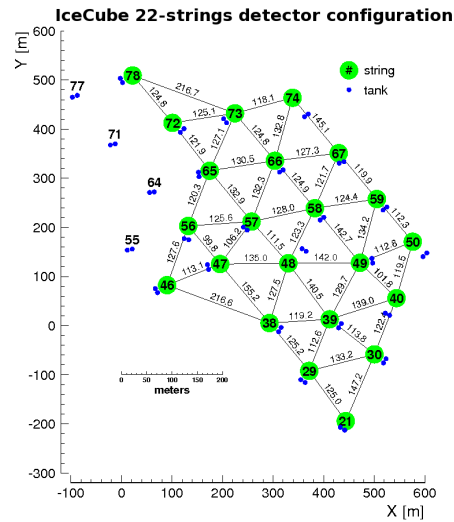}
 \end{center}
 \caption{The IceCube detector configuration. The filled green circles are the positions of IceCube 
strings and the filled blue circles display the position of the IceTop tanks.}
 \label{configs}
 \end{figure}

The remaining challenge for this analysis is accounting for the detector asymmetry, and unequal time coverage in the data due to the detector run selection. To illustrate the detector asymmetry Figure~\ref{configs} shows the IceCube 22 string geometrical configuration. This geometrical asymmetry results in a preferred reconstructed muon direction since the muons would pass by more strings in one direction in the detector compared to another. The combination of detector event asymmetry with a non-uniform time coverage would induce an azimuthal asymmetry and consequently artificial anisotropy of the arrival direction of cosmic rays. This asymmetry is corrected for by normalizing the azimuthal distribution.

\begin{figure}[h!]
 \begin{center}
 \noindent
 \includegraphics[width=\columnwidth]{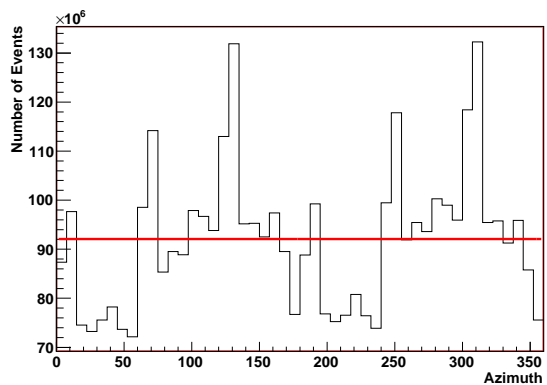}
 \end{center}
 \caption{The azimuthal distribution for the whole data set. This plot shows the number of 
events vs. the azimuth of the arrival direction of the primary CR particle. The horizontal red line
in the average number of events for the distribution.}
 \label{azimuth}
 \end{figure}

Figure~\ref{azimuth} shows the azimuthal distribution for the whole data set.  It displays the number of events vs. the azimuth of the arrival direction of the primary CR particle. Note that the asymmetry in the azimuthal distribution due to detector geometry is modeled well by simulation. 

To correct for the detector azimuthal asymmetry we apply an azimuthal normalization. The azimuthal distribution is parameterized by $N$, $n_i$, and $\bar{n}$, where $N$ is the total number of bins, $n_i$ is the number of events per bin and $\bar{n}$ is the average number of events, $\bar{n} = \frac{1}{N}\sum_{i=1}^{N} n_i$. $\bar{n}$ is denoted by the horizontal red line in Figure~\ref{azimuth}. The azimuthal normalization is applied by weighting each event by $\frac{\bar{n}}{n_i}$ for that event.  

In addition to the azimuthal asymmetry we also observe a non uniform zenith angle distribution (more events arrive from the zenith than from the horizon).  Due to this declination dependence, the sky is divided into four declination bands such that the data is approximately equally distributed among the declination bands. For each band the azimuthal distribution is normalized for the whole year.  The relative intensity for each bin in the 2-dimensional skymap is then calculated by dividing the number of events per bin by the total number of events per declination.


\section{Results:}

\begin{figure}[h!]
 \begin{center}
 \noindent
\includegraphics[width=\columnwidth]{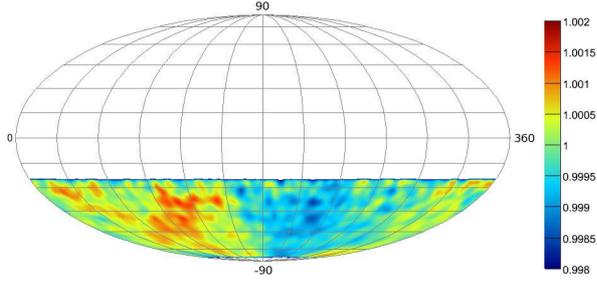} 
 \end{center}
 \caption{The IceCube skymap in equatorial coordinates (Declination (Dec) vs. Right Ascension (RA)). 
The color scale is the relative intensity. }
 \label{fig1}
 \end{figure}

\begin{figure}[h!]
 \begin{center}
 \noindent
 \includegraphics[width=\columnwidth]{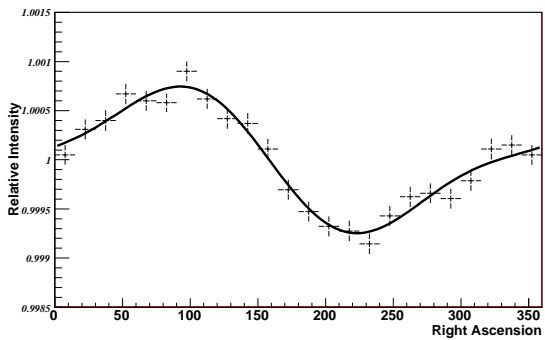} 
 \end{center}
 \caption{The 1-dimensional projection of the IceCube 2-dimensional skymap. The line is the first and second harmonic function fit.}
 \label{1d}
 \end{figure}

\begin{figure}[h!]
 \begin{center}
 \noindent
 \includegraphics[width=\columnwidth]{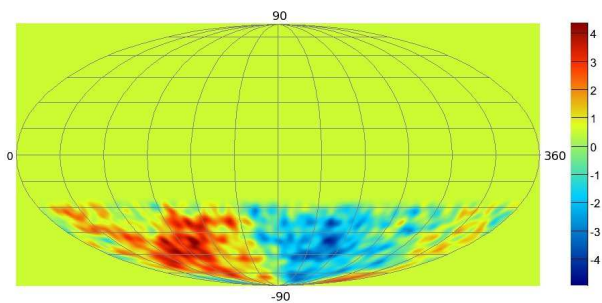} 
 \end{center}
 \caption{The IceCube significance skymap in equatorial coordinates (Dec vs. RA). The color scale is 
the significance.}
 \label{sig}
 \end{figure}

Figure~\ref{fig1} shows the southern hemisphere skymap for well reconstructed downward going muons for the IceCube 22-Strings data set.  The skymap is plotted in equatorial coordinates. The color scale represents the relative intensity of the rate for each bin per declination band where each bin rate is calculated by dividing the number of events for that bin over the average number of events for that bin's declination band. The plot shows a large scale anisotropy in the arrival direction of cosmic rays.  The amplitude and the phase of this anisotropy is determined by projecting the 2-dimensional skymap in Right Ascension (RA) as shown in Figure~\ref{1d}.  Figure~\ref{1d} shows the relative intensity vs. the RA. The data are shown in points with their error bars. The fit is the first and second-order harmonic function in the form of $\sum_{i=1}^{n=2}(A_{i} \times \cos(i~((RA)-\phi_{i}))) + B$ where $A_{i}$ is the amplitude and $\phi_{i}$ is the phase and $B$ is a constant. The fit $A_{i}$, $\phi_{i}$ and $\chi^{2}/ndof$ for the first and second harmonic fit are listed in table~\ref{tab1}. The significance of the 2-dimensional skymap is shown in Figure~\ref{sig}. The significance is calculated for each bin from the average number of events for that bin's declination band.  Note that the significance of several bins in the excess region is greater than $4 \sigma $ and in some bins in the deficit region is smaller the $ -4 \sigma$.

 \begin{table}[h!]

 \begin{center}
    \begin{tabular}{ |  p{1cm} | p{1.5cm} |  p{1cm} | p{1.5cm} |  p{1cm} | l |}
    \hline
    $A_{1}$. ($10^{-4}$) & $\phi_{1}$  & $A_{2}$. ($10^{-4}$) &  $\phi_{2}$  & $\chi^{2}/ndof$\\ \hline
    $ 6.4 \pm 0.2$ & $ 66.4^{\circ} \pm 2.6^{\circ}$ &  $ 2.1 \pm 0.3$ &  $ -65.6^{\circ} \pm 4^{\circ}$  &22/19 \\ \hline
   \hline 
 \end{tabular}
\end{center}
\caption{The first and second harmonic fit amplitude, phase, and $\chi^{2}/ndof$.}
 \label{tab1}
\end{table}

To check for the stability of the measured large scale anisotropy we performed a number of checks with the data set. One check was applied by dividing the data into two sets where one set contains sub-runs with an even index number and the other set contains sub-runs with an odd index number (a sub-run on average contains events collected for $\sim 20$ minutes at a time). Another check is applied by dividing the data into two sets each set contains half the sub-runs selected randomly. The results for both tests were consistent.

In addition, stability checks are applied to test for daily and seasonal variation effects. To test for the daily variation effect the data were divided in two sets: The first set contains sub-runs with rate values greater than the mean rate value for that sub-run's day. The second set contains sub-runs with rate values less than the mean rate value for that sub-run's day. Furthermore, to test for the seasonal variation effect the data set is divided in two sets: The first set holds the winter month's sub-runs (June-Oct.). The second set holds the summer month's sub-runs (Nov.-March). For both tests we see no significant changes in the value of the anisotropy of the two data sets.

\section{Energy Dependence:} 

\begin{figure}[h!]
 \begin{center}
 \noindent
 \includegraphics[width=\columnwidth]{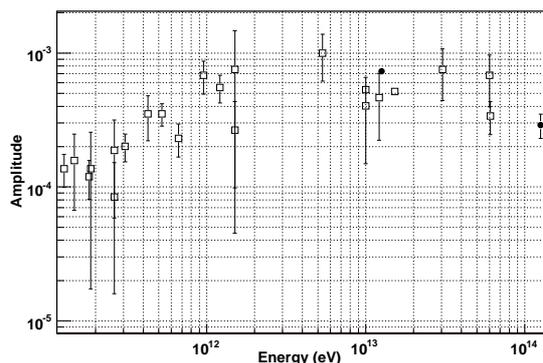} 
 \end{center}
 \caption{ The amplitude vs. energy in eV. The filled circles markers are the result of this analysis and the empty square markers are the result from previous experiments
(\cite{sk},~\cite{r1},~\cite{r2},~\cite{r3},~\cite{r4},~\cite{r5},~\cite{r6},~\cite{r7},~\cite{r8},~\cite{r9},~\cite{r10},
~\cite{r11},~\cite{r12},~\cite{r13},~\cite{r14},~\cite{r15},~\cite{r16},~\cite{r17}).}
 \label{fig2}
 \end{figure}

In order to better understand the possible nature of the anisotropy we have searched for energy dependent effects using our data. To determine the energy dependence for the signal we divided the data into two energy bins. To accomplish that and to ensure constant energy distribution along our sky both the number of sensors triggered by the event and the zenith angle of the event are used for the energy bands selection. The first energy bin contains $3.8 \times 10^{9}$ events with a median per CR particle of 12.6 TeV and $90 \%$ of the events between 2 and 158 TeV. The second energy bin contains  $9.6 \times 10^{8}$ events with a median energy per CR particle of 126 TeV and $90 \%$ of the events between 10 TeV and 1 PeV. Each 2-dimensional skymap is projected to 1-dimensional variations in RA. In comparison to previous experiments the 1-dimensional RA distribution is fitted to a first harmonic fit. The first harmonic fit amplitude and phase for the first energy band are $A_{1} = (7.3 \pm 0.3) \times 10^{-4}$ and $\phi_{1} = 63.4^{\circ} \pm 2.6^{\circ}$, while the amplitude and phase for the first energy band are $A_{1} = (2.9 \pm 0.6) \times 10^{-4}$ and $\phi_{1} = 93.2^{\circ} \pm 12^{\circ}$. Figure~\ref{fig2} shows the amplitude vs. energy of this analysis (in filled circles) in comparison to previous experiments (In empty squares). Note that the amplitude in this analysis shows a decrease of the harmonic amplitude value at the higher energies for the energy ranges of 10-100 TeV.  



\section{Conclusion}
In this analysis we present the first 2-dimensional skymap for the southern hemisphere of $4.3$ billion cosmic rays with a median angular resolution of $3^{\circ}$ and a median energy per CR particle of 14 TeV as observed by IceCube. A large cosmic ray anisotropy with a first harmonic vector amplitude of  $A_{1} = (6.4 \pm 0.2) \times 10^{-4}$ and a phase of $\phi_{1} = 66.4^{\circ} \pm 2.6 ^{\circ}$ is observed. The significance of some bins in the excess and the deficit regions were found to be $ >| 4\sigma| $. This anisotropy is an extension of previously measured large scale anisotropy at the northern hemisphere reported by multiple experiments (\cite{ta}-~\cite{milagro}).  

In addition, we report on the anisotropy energy dependence. We report the amplitude of the first harmonic vector of the anisotropy for the two energy bands. The first energy band with a median energy per CR particle of 12.6 TeV the amplitude is found to be  $A_{1} = (7.3 \pm 0.3) \times 10^{-4}$. The second energy band with a median energy per CR particle of 126 TeV the amplitude is found to be $A_{1} = (2.9 \pm 0.6) \times 10^{-4}$.  The amplitude energy dependence is found to follow a decreasing trend with energy.

\section{Acknowledgments}
We acknowledge the support from the following agencies: U.S. National Science Foundation-Oﬃce of Polar Program, U.S. National Science Foundation-Physics Division, University of Wisconsin Alumni Research Foundation, U.S. Department of Energy, and National Energy Research Scientiﬁc Computing Center, the Louisiana Optical Network Initiative (LONI) grid computing resources; Swedish Research Council, Swedish Polar Research Secretariat, and Knut and Alice Wallenberg Foundation, Sweden; German Ministry for Education and Research (BMBF), Deutsche Forschungsgemeinschaft (DFG), Germany; Fund for Scientiﬁc Research (FNRS-FWO), Flanders Institute to encourage scientiﬁc and technological research in industry (IWT), Belgian Federal Science Policy Oﬃce (Belspo); the Netherlands Organisation for Scientiﬁc Research (NWO); M. Ribordy acknowledges the support of the SNF (Switzerland); A. Kappes and A. Gro$\ss$ acknowledge support by the EU Marie Curie OIF Program.

\bibliographystyle{ieeetr}
\bibliography{icrc1340}

\end{document}